# Intervalley splittings of Si quantum wells

S.-H. Park[1], Y. Y. Lee[2], and Doyeol Ahn*[2]

[1]*Department of Electronics Engineering, Catholic University of Daegu, Hayang, Kyeongbuk 712-702, Republic of Korea*

[2]*Institute of Quantum Information Processing and Systems, University of Seoul, Seoul 130-743, Republic of Korea*

**Abstract**

Multi-valley effective mass theory for silicon quantum well structure is studied taking into account the external fields and the quantum interfaces. It is found that the phenomenological delta function potential, employed to explain the valley splitting caused by the quantum well interface in the previous work [Ref. 10], can be derived self-consistently from the multi-valley effective mass theory. Finite element method is used to solve the multi-valley effective equations. Theoretical predictions are in a reasonably good agreement with the recent experimental observation of valley splitting in a $SiO_2/Si/SiO_2$ quantum well, which prove the validity of our approach.

*To whom correspondence should be addressed.
E-mail: *dahn@uos.ac.kr* ; *davidahn@hitel.net*

Intervalley splitting of Si quantum structures is of current technological interest for the potential applications to future silicon hetero-structure devices that involve quantum effects, especially for the scalable quantum computation [1-11]. The lowest conduction band of a silicon crystal is known to have six equivalent minima of ellipsoidal shape called valleys along the [001] direction [12]. In the case of a strained quantum well or a quantum dot, the valley degeneracy is reduced to two fold [10]. It is found that the wave functions localized around different valleys become coupled in silicon quantum wells or quantum dots and form polarized intervalley states, which behave like pseudo spins [7, 9]. These polarized states offer an unexplored degree of freedom where the valley index specifying the valley in **k** space could be taken in analogy with the subband index of real-space quantized systems or Landau levels [5]. Recently, a single qubit and an elementary two-qubit quantum gate based on polarized intervalley states have been suggested by one of us [7]. Estimated decoherence time of these polarized states is in the order of micro-sec to milli-sec which is comparable to that of the spin qubit. The long decoherence time is especially important for the application in quantum computation. Existence of similar polarized intervalley states at zero magnetic fields was confirmed independently by Takashina et al. [9]. Also, controllable intervalley splitting in silicon quantum well has been studied as a potential source of decoherence in the spin-based silicon quantum computation scheme [10-11].

Considering the technological significance of the polarized intervalley states and valley splitting on emerging silicon quantum devices, it is crucial to investigate the origin of the valley splittings and to find an efficient but accurate method to calculate the quantum states for the control of device structures. Effective mass approximation (EMA) is known to be one of the most effective methods to treat the shallow donors in bulk silicon taking into account the valley states for the last several decades [13-19]. There has been renewed interest in EMA as a practical method to calculate the quantum states of silicon quantum structures [6-10]. Recently, it is argued that the closed effective mass description, which can not incorporate the microscopic details of the interface of the quantum structure, is believed to cause quantitative errors in the calculation [10]. In order to overcome this difficulty, Friesen et al. [10] incorporated a phenomenological delta function with a valley coupling parameter $v_v$ in EMA and was able to explain the oscillation of the valley splitting predicted by the tight-binding theory [6]. In this approach, the delta function which relates the quantum well interface, is supposed to be responsible for the valley splitting. However, an additional fitting parameter was needed to obtain the correspondence between the EMA and the tight-binding calculation.

In this paper, by extending the effective mass theory of Shindo [17] and Ohkawa [18] to the silicon quantum structure, we first show that the phenomenological delta function term proposed by Friesen et al. [10] arises from the multi-valley EMA, self-consistently. Finite element method is used to calculate the quantum states using multi-valley EMA. It turns out that the calculated valley splitting oscillation in our model agrees with that of the tight-binding theory both qualitatively and quantitatively. We compared our theoretical results with recent experimental observation of valley splittings in a $SiO_2/Si/SiO_2$ quantum well and found a reasonably good agreement.

We consider a Si-SiGe quantum well with the z-direction assumed to be along the Si (001) surface. Our model can also be extended to the cases of a $Si-SiO_2$ quantum well and a Si quantum dot. Based on Kohn-Luttinger effective mass theory [13-19], the envelope function for the quantum state in a Si quantum well is given by $F(\vec{r}) = \sum_{\vec{k}} F(\vec{k}) \exp(i\vec{k} \cdot \vec{r})$ for $F(\vec{k}) = \sum_{i} \alpha_i F_i(\vec{k})$, where $F_i(\vec{k})$ is centered about the $i$th minimum. The constants $\alpha_i$ can be determined from the group theoretical considerations [20]. The equation of motion for $F_i(\vec{k})$ becomes [7]

$$\varepsilon_i(\vec{k})F_i(\vec{k}) + \sum_j \sum_{\vec{k}'} D^{ij}_{\vec{k},\vec{k}'} V(\vec{k}-\vec{k}')F_j(\vec{k}') = \varepsilon F_i(\vec{k}), \tag{1}$$

where $\varepsilon_i(\vec{k})$ is the energy dispersion relation of the $i$-th valley, $V(\vec{k})$ the Fourier component of the total potential, and $D^{ij}_{kk'}$ is the inter-valley coupling term which can be derived from the cell periodic function for the conduction band as

$$\begin{aligned} D^{ij}_{kk'} &= D^{ij}_{\vec{K}_i+\vec{\kappa},\vec{K}_j+\vec{\kappa}} \\ &\cong D^{ij}_{\vec{K}_i,\vec{K}_j} + \vec{\kappa}\cdot\frac{\partial}{\partial \vec{K}_i}D^{ij}_{\vec{K}_i,\vec{K}_j} + \vec{\kappa}\cdot\frac{\partial}{\partial \vec{K}_j}D^{ij}_{\vec{K}_i,\vec{K}_j} \\ &= I_{ij} + \vec{\kappa}\cdot\vec{J}_{ij} + \vec{\kappa}\cdot\vec{J}_{ij}, \end{aligned} \tag{2}$$

where $\vec{K}_i$ is the wave vector at the minimum at the $i$-th valley. There have been two types of EMA. One approach [15-16] is based on the method of Fritzsche [21] and Twose [22] and the other is the multi-valley effective mass theory of Shindo [17] and Ohkawa [18]. The major difference is the neglect of the coupling of Bloch functions in different bands in the former. Within the frame of multi-valley effective mass theory [17, 18], the equation of motion for $F_l(\vec{r}) = \sum_{\vec{k}} F_l(\vec{k})\exp(i\vec{k}\cdot\vec{r})$ can be written down as [7]

$$\left[H_l(k_x,k_y,-i\partial/\partial z) + V(\vec{r}) - E\right]F_l(\vec{r}) + \sum_{l'\neq l} H_{ll'}(\vec{r},-i\vec{\nabla})F_{l'}(\vec{r}) = 0. \tag{3}$$

Here,

$$H_l(-i\vec{\nabla}) = \frac{\hbar^2 k_x^2}{2m_x} + \frac{\hbar^2 k_y^2}{2m_y} - \frac{\hbar^2}{2m_z}\frac{\partial^2}{\partial z^2} + \frac{e\hbar B}{2m_x}yk_x + \frac{e\hbar B}{2m_y}xk_y + \frac{e^2 B^2}{8}\left(\frac{x^2}{m_y} + \frac{y^2}{m_x}\right), \tag{4}$$

and

$$H_{ll'}(\vec{r},-i\vec{\nabla})$$
$$= I_{ll'} \exp[-i(\vec{K}_l - \vec{K}_{l'})\cdot\vec{r}](V(\vec{r}))$$
$$-i(\vec{J}_{ll'} \cdot \vec{\nabla})\exp[-i(\vec{K}_l - \vec{K}_{l'})\cdot\vec{r}](V(\vec{r})) \quad (5)$$
$$+ \exp[-i(\vec{K}_l - \vec{K}_{l'})\cdot\vec{r}](V(\vec{r}))(-i\vec{J}'_{ll'} \cdot \vec{\nabla})$$

and $\quad V(\vec{r}) = V_c(\vec{r}) + e\vec{F}\cdot\vec{r}$, $\quad\quad\quad\quad\quad\quad\quad\quad\quad\quad\quad\quad$ (6)

where $m_x, m_y, m_z$ are effective masses along x, y, z directions in each valley, $E$ is quantized energy, $\vec{K}_l$ is the wave vector at the minimum at the $l$-th valley, $I_{ll'}, \vec{J}_{ll'}, \vec{J}'_{ll'}$ are inter-valley coupling terms, $V_c(\vec{r})$ is the confinement potential, and $\vec{F}$ is an applied electric field.

In order to calculate the inter-valley coupling terms, we assume that $D^{ll'}_{\vec{K}_l,\vec{K}_{l'}}$ can be expressed by the following simple form, $D^{ll'}_{\vec{K}_l,\vec{K}_{l'}} = \alpha \vec{e}_l \cdot \vec{e}_{l'} + \beta$, where $\vec{e}_l$ is the unit vector in the direction of $l$-th axis and $\alpha, \beta$ are constants to be determined from the band-structure parameters. In this paper, we use the value of $\alpha = 0.6086$ and $\beta = 0.3915$ [7]. Then from equation (2), we obtain

$$I_{ll'} = \frac{1}{2}(1+\vec{e}_l \cdot \vec{e}_{l'}) - \frac{1}{2}(1-\vec{e}_l \cdot \vec{e}_{l'})\cos(2\lambda_K), \quad (7)$$

and

$$\vec{J}_{ll} = \vec{e}_l(1-\vec{e}_l \cdot \vec{e}_l)\frac{\partial \lambda_K}{\partial K}\sin(2\lambda_K), \quad (8)$$

with $\quad \tan(2\lambda_K) = \dfrac{2TK}{\varepsilon_G}$, where T=1.08 a.u. and $\varepsilon_G = 0.268 Ry$. The inter-valley coupling between the valley 5 and the valley 6 (along z-axis) is approximated by

$$H_{56} = -I_{56}\exp[-i(\vec{K}_5 - \vec{K}_6)\cdot\vec{r}](V_c(\vec{r}) + eFz)$$
$$- i|J_{56}|\frac{\partial}{\partial z}\left[\exp[-i(\vec{K}_5 - \vec{K}_6)\cdot\vec{r}](V_c(\vec{r}) + eFz)\right], \quad (9)$$

with

$$I_{56} = -\cos(2\lambda_K) = -0.217,$$
$$J_{56} = J'_{56} = 2\frac{\partial \lambda_K}{\partial K}\sin(2\lambda_K) = \frac{0.414}{K}. \quad (10)$$

If we substitute equations (9) and (10) into equation (5) for $l=5$ and $l'=6$ and assume that the electric field $F$ is in the z-direction, we obtain the valley splitting in the following form:

$$\Delta(F) \approx 2\left|\int d\vec{r} \exp(-2iK_o z)|\Psi_0(\vec{r})|^2 \left(1.045\, V(\vec{r}) + \frac{0.414}{K_0}\frac{\partial V(\vec{r})}{\partial z}\right)\right|, \quad (11)$$

where $K_0 = 0.85 \times 2\pi/a$, $a$ is the silicon lattice constant, $\Psi_0$ is the ground state of a single valley, and $V(\vec{r}) = V_C(\vec{r}) + eFz$ with $V_C$ the confinement potential and $F$ the applied electric field. Here $\Delta(F)$ is the off-diagonal element of the Hamiltonian (3). From equations (9) and (11), we obtain

$$\frac{\partial V_C(\vec{r})}{\partial z} = \Delta E_C \delta(z - z_i), \quad (12)$$

and, as a result,

$$\frac{\partial V(\vec{r})}{\partial z} = \Delta E_c \delta(z - z_i) + eF, \quad (13)$$

where $\Delta E_C$ is the conduction band discontinuity and $z_i$ is the quantum well interface. Equations (11) to (13) indicate that the valley splitting is caused by both the applied electric field and the quantum well interface. On the other hand, Friesen et al. [10] assumed that the valley splitting is caused only by the delta functions near the interface. It is interesting to compare the phenomenological effective coupling constant $V_v = 7.2 \times 10^{-11}$ of reference (10) with our result, $\frac{0.414}{K_o} = 4.42 \times 10^{-11}$ which is obtained from the first principle theory. In the previous work, we did not include the delta function term (12) in the calculation and, as a result, we obtained an underestimation of the valley splittings.

In Fig. 1, we show the valley splittings of a $Si_{0.7}Ge_{0.3}/Si/Si_{0.7}Ge_{0.3}$ quantum well as functions of the well with (in angstrom unit) at zero electric field in order to compare our results with other approaches. Our results agree well with the tight-binding calculations shown in the Fig. 3 of reference 10, both qualitatively and quantitatively. On the other hand, an additional fitting parameter was need to match the EMA results of reference 10, which is based the approach of Fritzsche [21] and Twose [22], with the tight-binding calculations.

Fig. 2 shows the effects of off-diagonal Hamiltonian (11) on the subband energy levels of $SiO_2/Si/SiO_2$ quantum well as a function of the well-width with an applied electric field of $1 \times 10^7 V/m$ by comparing the case (a) without the off-diagonal element and (b) with off-diagonal element when we solve multi-valley effective mass equations (3). We assumed barrier width of 6 nm in the computation. In Fig. 2 (b), one can see that the valley splittings show the oscillatory behavior but it decrease rapidly with the increasing well width. This can be explained by examining the integrand of equation (11). As the well with increase the delta function peak occurs for large value of |z| but the wave function decreases very rapidly together with the fast oscillating exponential term for the large value of z in eq. (11). As the result, the valley splitting either saturates or decreases with increasing quantum well width.

In Fig. 3, we show the effects of an applied electric field *F* on the subband energy levels of 6 nm $SiO_2/Si/SiO_2$ quantum well. We assumed the barrier width of 6 nm in the computation. At zero electric field, the intervalley splitting is caused by the quantum well potential and its first order derivatives at the interface and is dominated by the delta function term described in eq. (12). For large electric field, the valley splitting is mainly due to the applied electric field $F$ as can be seen from Fig. 3.

In Fig. 4, we show the valley splittings of a $SiO_2/Si/SiO_2$ quantum well as functions of the well with (in angstrom unit), an applied electric field $F$, and subband indices in order to compare our results with recent experimental results [9]. Takashina et al. [9] observed the valley splitting of tens of meV for a 8 nm wide Si quantum well between oxide layers for the first time. Previous observations showed the valley splittings are under a meV range. The range of the bias $V_{BG}$ can be calibrated to the applied electric field $F$ between $2.15 \times 10^7 \, V/m$ to $1.5 \times 10^8 \, V/m$. In Ref. 9, the valley splitting of 23 meV was obtained for the case of $V_{BG} = 60 \, V$ or $F = 1.29 \times 10^8 \, V/m$. On the other hand, Fig. 4(c) shows that the valley splitting is around 15 meV when the field is $1 \times 10^8 \, V/m$ and we can see that the agreement of the experimental results with theory is reasonable. For the bulk Si inversion layer, the valley splitting $\Delta$ (in meV unit) is estimated by the experimental formula, $\Delta = 1.14 n_s$ where the surface charge density $n_s$ in unit of $10^{12} cm^{-2}$ [12] or $\Delta = 0.718 F$ for the electric field in unit of $10^7 V m^{-1}$. Our result shows that the splitting is larger than the bulk Si inversion layer by a factor of 2 to 3 due to presence of the quantum well interface.

In this summary, we study the multi-valley effective mass theory for silicon quantum structures which can take into account the external field and the quantum interface. Finite element method is used to calculate the polarized states. It is found that the phenomenological delta function potential, employed to explain the valley splitting caused by the quantum well interface in the work of Friesen et al. [10], can be derived self-consistently from the first principle theory. We found a reasonably good agreement of our theoretical predictions with the recent experimental observation [9] of valley splitting in $SiO_2/Si/SiO_2$, which prove the validity of our approach.

**Acknowledgements**

This work was supported by the Korea Science and Engineering Foundation, the Korean Ministry of Science and Technology through the Accelerated Research Initiatives Program under the contract no. R17-2007-010-01001-0(2007).

**References**

1. P. Weitz, R. J. Haug, K. von Klitzing, and F. Schäffer, *Surf. Sci.* **361**, 542 (1996).
2. S. J. Koester, K. Ismail, and J. O. Chu, *Semicond. Sci. Technol.* **12**, 384 (1997)
3. V. S. Khrapai, A. A. Shashkin, and V. T. Dolgopolov, *Phys. Rev. B* **67**, 113305 (2003).
4. K. Lai, W. Pan, D. C. Tsui, S. Lyon, M. Mühlberger, and F. Schäffer, *Phys. Rev. Lett.* **93**, 156805 (2004).
5. K. Takashina, A. Fujiwara, S. Horiguchi, Y Takahashi and Y. Hirayama, *Phys. Rev. B* **69**, 161301(R) (2004).
6. T. B. Boykin, G. Klimeck, M. A. Eriksson, M. Friesen, S. N. Coppersmith, P. von Allmen, F. Oyafuso and S. Lee, *Appl. Phys. Lett.* **84**, 115 (2004).
7. D. Ahn, *J. Appl. Phys.* **98**, 033709 (2005).
8. M. Friesen, *Phys. Rev. Lett.* **94**, 186403 (2005).
9. K. Takashina, A. Fujiwara, Y. Takahashi, Y. Hirayama, *Phys. Rev. Lett.* **96**, 236801 (2006).
10. M. Friesen, S. Chutia, C. Tahan and S. N. Coppersmith, *Phys. Rev. B* **75**, 115318 (2007).
11. S. Goswami, K. A. Slinker, M. Friesen, L. M. Mcguire, J. L. Truitt, C. Tahan, L. J. Klein, J. O. Chu, P. M. Mooney, D. W. Van der Weide, R. Joynt, S. N. Coppersmith and M. A. Eriksson, *Nature Phys.* **3**. 41 (2007).
12. T. Ando, A. B. Fowler and F. Stern, *Rev. Mod. Phys.* **54**, 437 (1982).
13. J. M. Luttinger and W. Kohn, *Phys. Rev.* **97**, 869 (1955).
14. W. Kohn and J. M. Luttinger, *Phys. Rev.* **98**, 915 (1955).
15. T. H. Ning and C. T. Sah, *Phys. Rev. B.* **4**, 3468 (1971).
16. S. T. Pantelides and C. T. Sah, *Phys. Rev. B* **10**, 621 (1974).
17. K. Shindo and H. Nara, *J. Phys. Soc. Japan* **40**, 1640 (1976).
18. F. J. Ohkawa, *J. Phys. Soc. Japan* **46**, 736 (1979).
19. T. Ando, *Phys. Rev. B* **19**, 3089 (1979).

20. F. A. Cotton, *Chemical Applications of Group Theory* (John Wiley & Sons, New York, 1990).
21. H. Fritzsche, *Phys. Rev.* **125**, 1560 (1962)
22. W. D. Twose, in the Appendix of Ref. 21.

**Figure Captions**.

**Fig. 1** We show the valley splittings of a $Si_{0.7}Ge_{0.3}/Si/Si_{0.7}Ge_{0.3}$ quantum well for the ground state with zero applied electric field. Our results agree well with the tight-binding calculations shown in the Fig. 3 of reference 10, both qualitatively and quantitatively.

**Fig. 2** We show the effects of off-diagonal Hamiltonian (11) on the subband energy levels of $SiO_2/Si/SiO_2$ quantum well as a function of the well width with an applied electric field of $1 \times 10^7 V/m$ by comparing the case (a) without the off-diagonal element and (b) with off-diagonal element when we solve multi-valley effective mass equations (3). We assumed barrier width of 6 nm in the computation.

**Fig. 3** We show the effects of an applied electric field F on the subband energy levels of 6 nm $SiO_2/Si/SiO_2$ quantum well. We assumed the barrier width of 6 nm in the computation.

**Fig. 4** Valley splittings of a $SiO_2/Si/SiO_2$ quantum well as functions of the well with (in angstrom unit), an applied electric field $F$, and subband indices are shown. The calculated valley splitting of 29.9 meV when the field is $1 \times 10^8 V/m$ can be compared with the experimental splitting [9] of 23 meV for the case of $V_{BG} = 60 V$ or $F = 1.29 \times 10^8 V/m$, favorably.

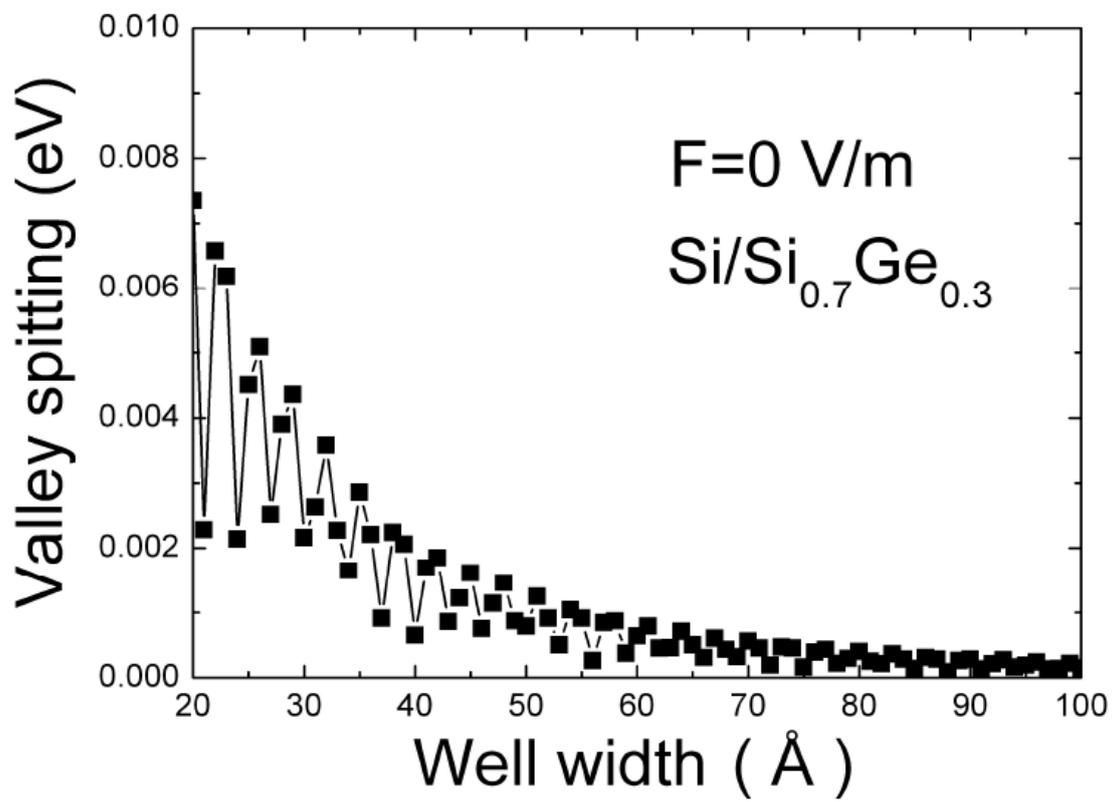

Fig. 1

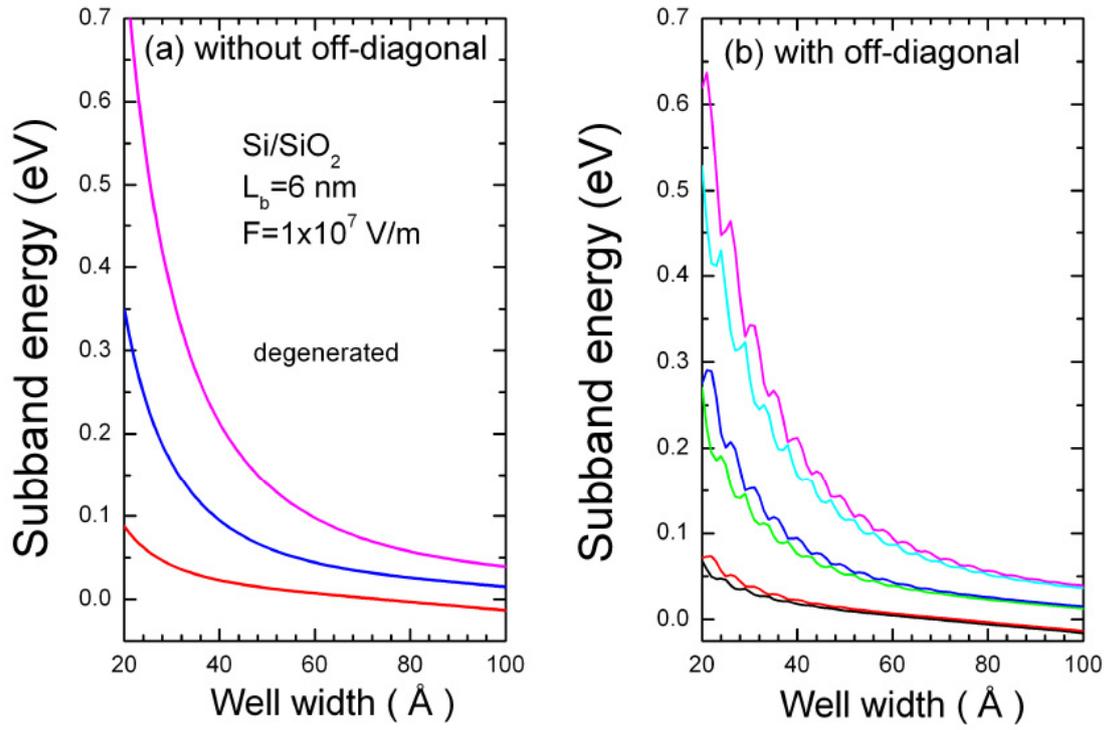

Fig. 2

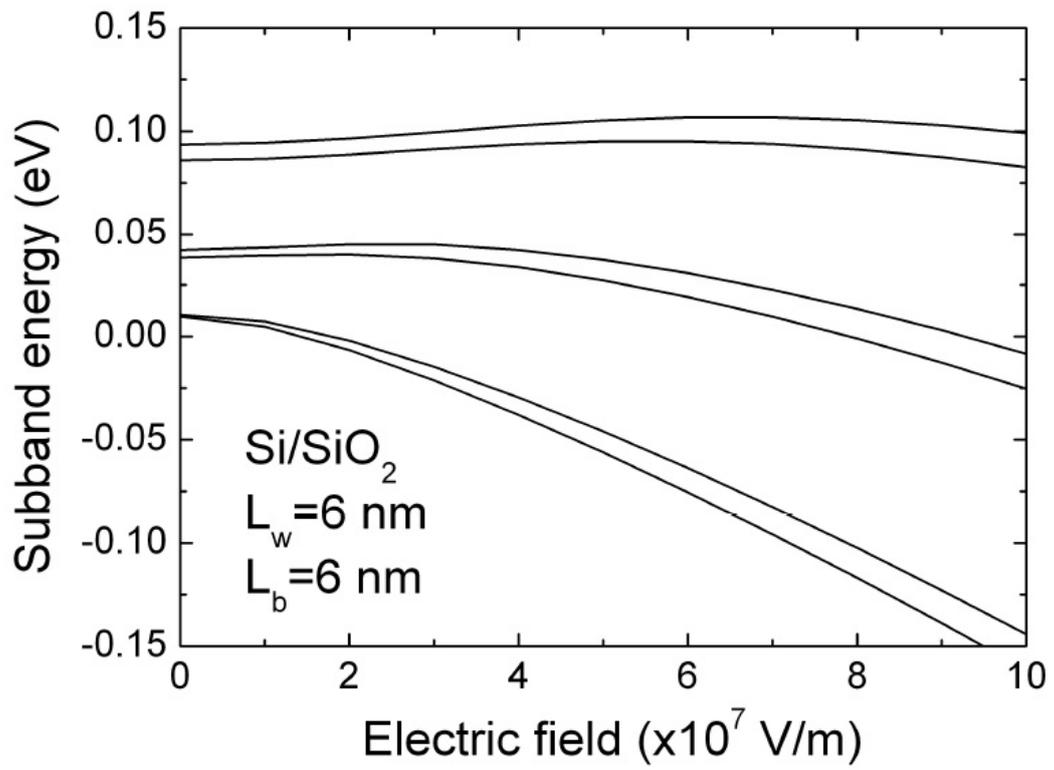

Fig. 3

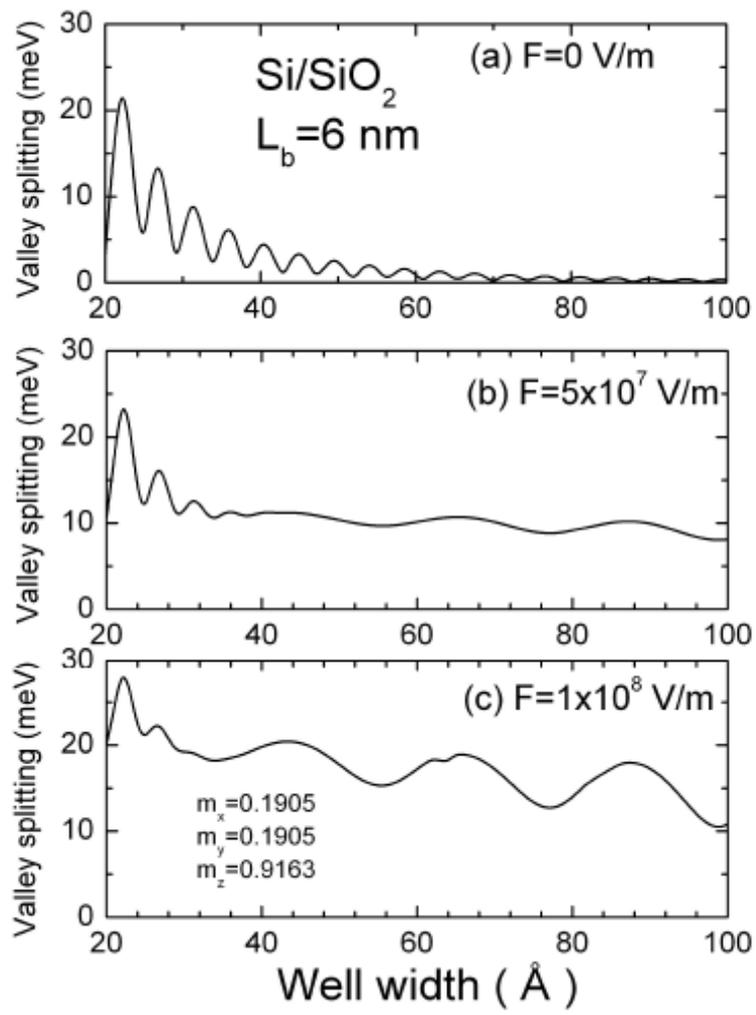

Fig. 4